\documentclass{llncs}

\usepackage{amsfonts,amsmath}
\usepackage{graphics}
\usepackage{float}
\usepackage{epsfig}
\usepackage{subfigure}
\newcommand{\remove}[1]{}

\title{Social determinants of content selection in the age of (mis)information}

\author{Alessandro Bessi\inst{1,2} \and Guido Caldarelli \inst{2,3} \and Michela Del Vicario\inst{2} \and Antonio Scala \inst{3} 
\and Walter Quattrociocchi\inst{2,4}}
\institute{IUSS Institute for Advanced Study, Piazza della Vittoria 5, 27100 Pavia, Italy
\and
 IMT Alti Studi Lucca, piazza S. Ponziano 6, 55100 Lucca, Italy 
\and
ISC-CNR Uos "Sapienza", 00185 Roma, Italy
\and
 Laboratory for the modeling of biological and socio-technical systems, Northeastern University, Boston, MA 02115 USA}

\begin{document}

\maketitle
\begin{abstract}

Despite the enthusiastic rhetoric about the so called \emph{collective intelligence}, conspiracy theories -- e.g. global warming induced by chemtrails or the link between vaccines and autism -- find on the Web a natural medium for their dissemination. 
Users preferentially consume information according to their system of beliefs and the strife within  users of opposite narratives may result in heated debates.
In this work we provide a genuine example of information consumption from a sample of 1.2 million of Facebook Italian users. We show by means of a thorough quantitative analysis that information supporting different worldviews  -- i.e. scientific and conspiracist news -- are consumed in a comparable way by their respective users. Moreover, we measure the effect of the exposure to 4709 evidently false information (satirical version of conspiracy theses) and to 4502 debunking memes (information aiming at contrasting unsubstantiated rumors) of the most polarized users of conspiracy claims. We find that either contrasting or teasing consumers of conspiracy narratives increases their probability to interact again with unsubstantiated rumors.

\noindent\textbf{Keywords:} misinformation, collective narratives, crowd dynamics, information spreading.
\end{abstract}

\section{Introduction}
The large availability of data from online social networks (OSN) allows for the study of mass social dynamics at an unprecedented level of resolution. 
Along this path, recent studies have pointed out several important results in the emerging field of computational social science \cite{Lazer09,conte2012manifesto} ranging from the influence-based contagion, up to the emotional contagion, passing through the virality of false claims  \cite{kramer2014,Aral2014,Mocanu2014}. In particular in \cite{Mocanu2014,Bessi2014} it has been shown that massive digital misinformation permeates online social dynamics. 
Social interaction, healthcare activity, political engagement and economic decision-making are influenced by digital hyper-connectivity -- i.e. the increasing and exponential rate at which people, processes and data are connected and interdependent \cite{Lotan11,Lewis2012,Leskovec2010,kleinberg2013analysis,Richard2004,howard11-1,Moreno2011,Bond2012,Quattrociocchi2014,Aral2012}. Everyone can produce and access a variety of information actively participating in the diffusion and reinforcement of narratives. Such a process has been  dubbed as \emph{collective intelligence} \cite{Levy2000,Malone2007}. However, despite the enthusiastic rhetoric about the ways in which digital technologies have burst the interest in debating political or social relevant issues, their role in enforcing informed debates and shaping the public opinion still remain unclear. Indeed, the World Economic Forum listed \emph{massive digital misinformation} as one of the main risks for the modern society \cite{Davos13}. Conspiracy theories as alternative explanations to complex phenomena (e.g., globalization or climate change) find on the Web a natural medium for their dissemination and, not rarely, they are used as argumentation for policy making and foment collective debates \cite{itais2014}. 
Conspiracy theses tend to reduce the complexity of reality by explaining significant social or political aspects as plots conceived by powerful individuals or organizations. Since these kinds of arguments can sometimes involve the rejection of science, alternative explanations are invoked to replace the scientific evidence. For instance, people who reject the link between HIV and AIDS generally believe that AIDS was created by the U.S. Government to control the African American population \cite{Sunstein12}. 
The spread of misinformation in such trusted networks can be particularly difficult to detect and correct because of the social reinforcement -- i.e. people are more likely to trust an information originating from within their network or someway consistent with their system of beliefs \cite{mckelvey2013truthy,Meade2002,Mann2000,Garrett2013,Buckingham2012,Carletti2006,Centola2010,eRep,QuattrociocchiCL11,QuattrociocchiPC09,Bekkers2011,Quattrociocchi2014,Quattrociocchi2012}.
Since unsubstantiated claims are proliferating over the Internet, what would happen if they were used as the basis for policy making? Such a scenario makes crucial the quantitative understanding of the social determinants related to content selection, information consumption, and beliefs formation and revision.
Misinformation is pervasive and as a first reaction we noticed the emergence of blogs and pages devoted to debunk false claims, namely {\em hoaxbusters}. Meanwhile, the strong polarization of users with respect to one or another narrative (fomented by the possibility to ban and to write negative comments) triggered the proliferation of satirical pages  producing demential imitation of conspiracy theses (e.g., chemtrails containing sildenafil citratum -- i.e. the active ingredient of Viagra\texttrademark -- or the political action committee to abolish the thermodynamic laws), namely {\em trolls}.
In this work we provide a genuine example of robust generative patterns about information consumption on the Italian Facebook on a sample of 1.2 million of individuals. In particular, we show, through a thorough quantitative analysis, similar consumption patterns of information supporting different (and opposite) worldviews.
Then, we measure the social response of polarized users of alternative news to 4709 satirical  version of conspiracy theses and to 4502 debunking memes (information aiming at correcting the diffusion of unsubstantiated claims) for increasing level of user commitment on the preferred narratives (scientific news and conspiracy news). 
We find that, for polarized users of conspiracy-like claims the exposure to either debunking or parody of conspiracy claims, the survival probability -- i.e. the probability to continue consuming posts related to conspiracy -- increases with the user's commitment in the narrative. 

\section{Data Collection}

We want to characterize information consumption patterns of users with respect to the different and heterogeneous information belonging to different narratives.
In order to define the space of our investigation, we were helped by Facebook groups very active in the debunking of conspiracy theses (see acknowledgments section).
The resulting dataset is composed of 73 public Facebook pages divided in scientific news and conspiracist news for which we downloaded all the posts (and their respective users interactions) in a timespan of 4 years (2010 to 2014).
Furthermore, we consider also 6 pages very active in debunking conspiracy information, namely hoax-busters, and 2 pages satirizing conspiracy theories by diffusing intentional false information as a satirical imitation of conspiracy theses. These latter have produced information that went viral despite their evident satirical taste. Among these, the OGM yellow tomatoes and the violet carrots created by industries to satisfy aesthetic needs (notice that the first tomatoes arrived in Europe were yellow) or the wonderful anti-hypnotic effects of lemon (such a post received more than 45.000 shares).
The entire data collection process is performed exclusively with the Facebook Graph API \cite{fb_graph_api}, which is publicly available and which can be used through one's personal Facebook user account.
The exact breakdown of the data is presented in Table \ref{tab:data_dim}. 
The first category includes all pages diffusing conspiracist information -- pages which disseminate controversial information, most often lacking supporting evidence and sometimes contradictory of the official news. 
The second category is that of scientific dissemination including scientific institutions and scientific press having the main mission to diffuse scientific knowledge. 
We focus our analysis on the interaction of users with the public posts -- i.e. likes, shares, and comments.  Each of these actions has a particular meaning. A {\em like} stands for a positive feedback to the post; a {\em share} expresses the will to increase the visibility of a given information; and {\em comment} is the way in which online collective debates take form. Comments may contain negative or positive feedbacks with respect to the post.

\begin{center}
\begin{table}[H]
\centering
 \begingroup
\begin{tabular}{l|c|c|c|c|c|c}
\bf {  }  & \bf {Total} & \bf {Science} & \bf {Conspiracy}  & \bf {Hoaxbusters}  & \bf {Troll} \\ \hline 
Pages & $ 81 $ & $ 34 $ & $ 39 $  & $ 6 $ & $ 2 $\\
Posts & $ 271,296 $ & $ 62,705 $ & $ 208,591 $ & $ 4,502 $ & $ 4,709 $ \\
Likes & $ 9,164,781 $ & $ 2,505,399 $ & $ 6,659,382$ & $ 67,324 $& $ 40,341 $ \\
Comments & $1,017,509 $ & $ 180,918 $ & $ 836,591 $ & $ 17,883 $& $ 58,686 $\\
Unique Comments & $ 279,972 $ & $ 53,438 $ & $ 226,534 $& $ 5,115 $& $ 42,910 $\\
Unique Likes & $ 1,196,404 $ & $ 332,357 $ & $ 864,047 $& $ 12,427 $& $ 16,833 $\\
\end{tabular}\newline
 \endgroup
  \caption{ \textbf{Breakdown of Facebook dataset.} The number of pages, posts, comments and likes for all category of pages.}
  \label{tab:data_dim}
\end{table}
\end{center}

\section{Results and Discussion}
\subsection{On the fruition of distinct narratives}
\begin{figure}[H]
 \centering
{\includegraphics[width=.4\textwidth]{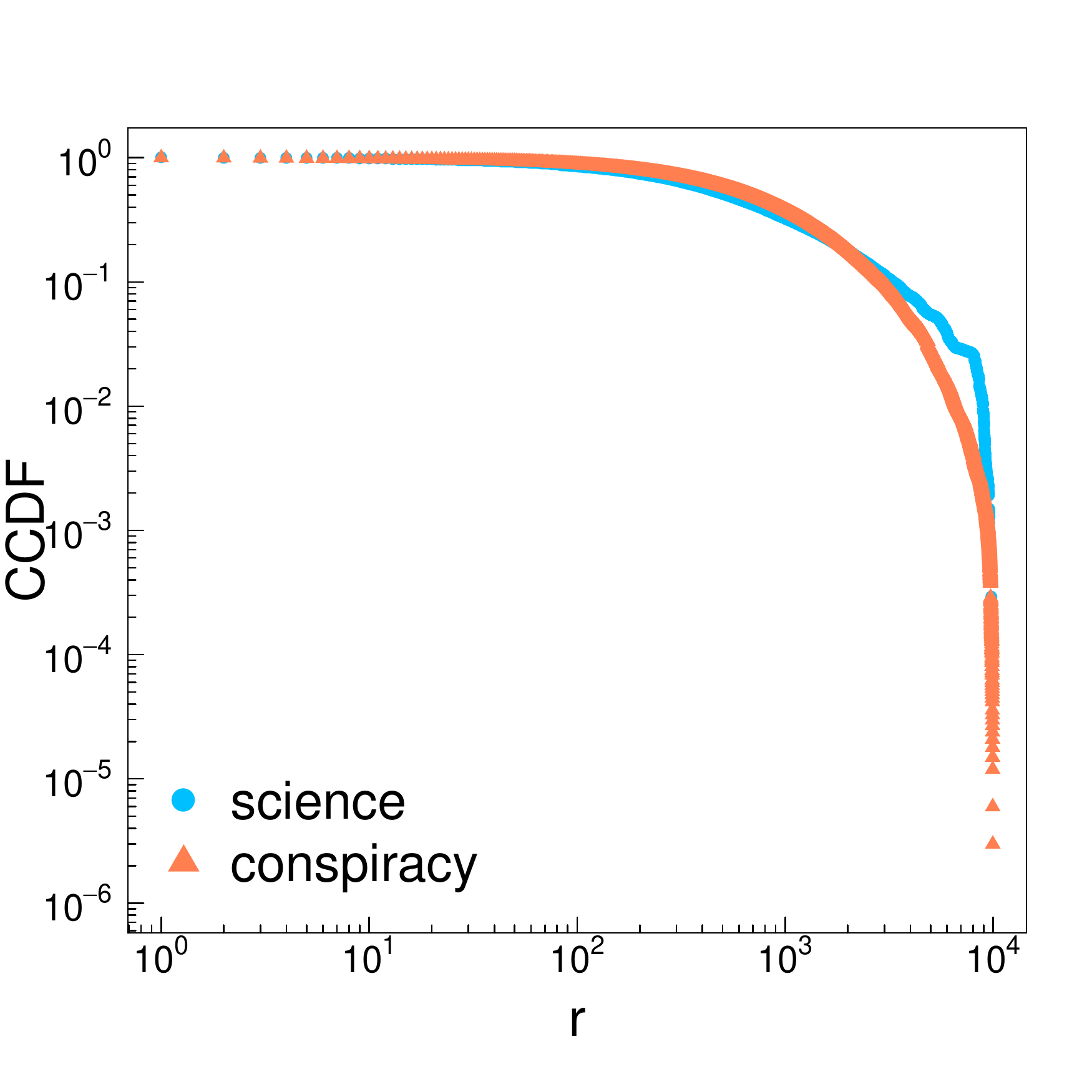}} 
\caption{\textbf{Fruition Patterns}: Empirical CCDF of the mean time interval (in hours) between likes for each user.  The two distributions are indicating a similar behavior in the rate of persistence of the users.}
\label{fig:fruition}
\end{figure}

We start our analysis by characterizing information consumption patterns by focusing on the behavior of usual consumers of conspiracy and scientific news.
Through a thresholding strategy we select the most active users in a specific category according to their liking activity on posts. 
As we assume \emph{likes} to be positive feedbacks with respect to the information reported on the post \cite{Viswanath:2009}, a user is labeled as polarized in one category if the 95\% of his likes is given on posts published on pages of such a category. Through this classification algorithm we are able to label $255,225$ users polarized in science and $790,899$ users polarized in conspiracy.
As a first level of approximation of the users interaction patterns we focus on the temporal dimension of the users persistence on one or the other category.
In Figure \ref{fig:fruition} we show the empirical complementary cumulative distribution function (CCDF) of the users' persistence rate, namely $r$, intended as the mean time interval (in hours) between likes of a user on posts of their preferred narrative. 
Mean and median for Science are, respectively, 1212 and 513 hours. Mean and median for Conspiracy are, respectively, 1155 and 665 hours.
Usual consumers of conspiracy and scientific news present a very similar interaction with the posts.

\begin{figure}[H]
 \centering
{\includegraphics[width=.4\textwidth]{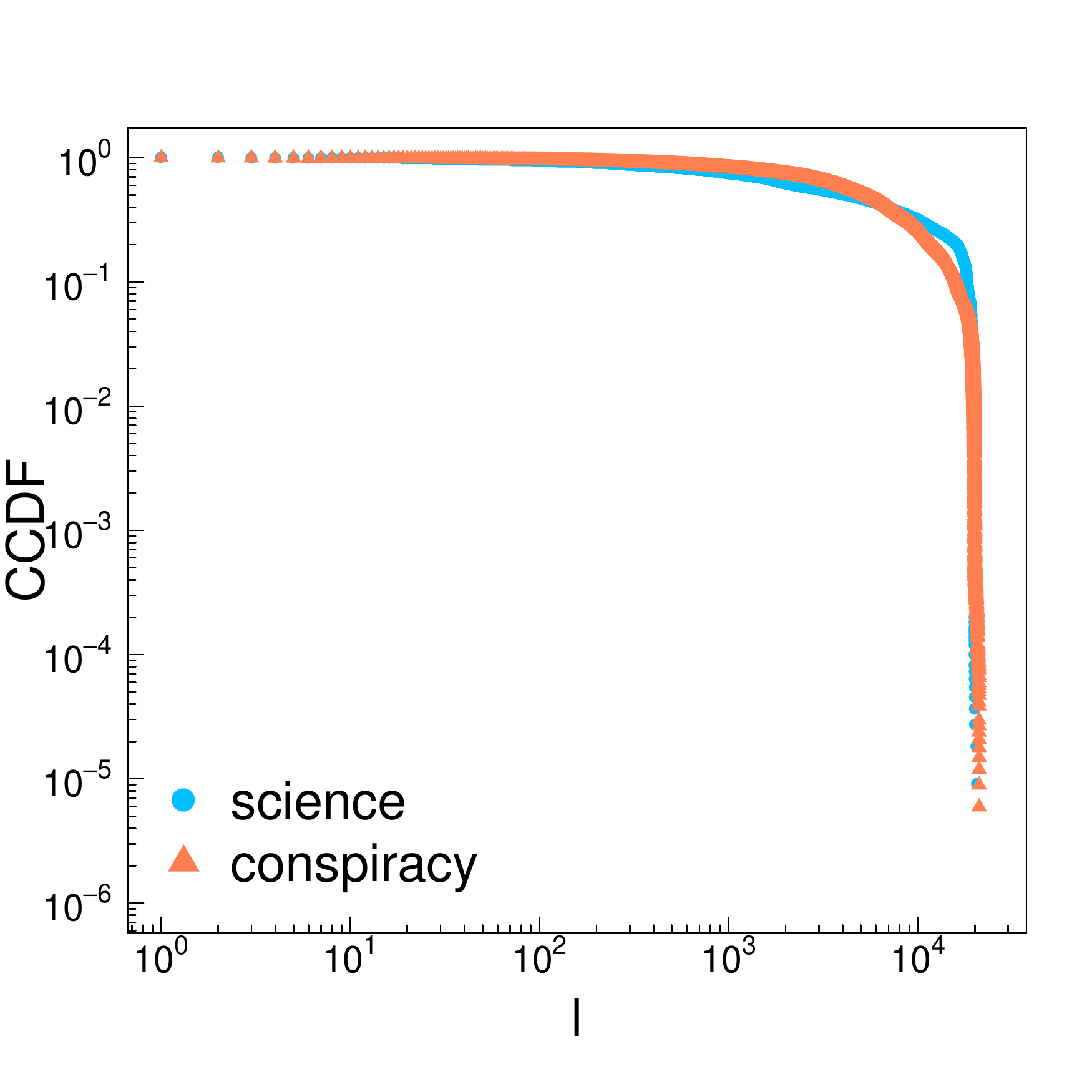}} 
\caption{\textbf{User's lifetime}. Empirical CCDF of user's lifetime (in hours) -- i.e. the time interval between the first and the last like of each polarized user in the category which he belongs to.}
\label{fig:survival}
\end{figure}

We zoom in at the level of the users' lifetime in the category on which they are assigned to. In Figure \ref{fig:survival} it is shown  the CCDF of users' lifetime, namely $l$ -- i.e., the time interval (in hours) between the first and the last like of the users on posts of the category which they are assigned to.
Mean and median user's lifetime for Science are, respectively, 6976 and 4126 hours; mean and median user's lifetime for Conspiracy are, respectively, 6651 and 5161 hours.

These results show that information belonging to different narratives are consumed in a similar way by they respective users.

\subsection{Engagement in narratives and responsivity to external information}

\begin{figure}[H]
 \centering
       \includegraphics[width=0.4\textwidth]{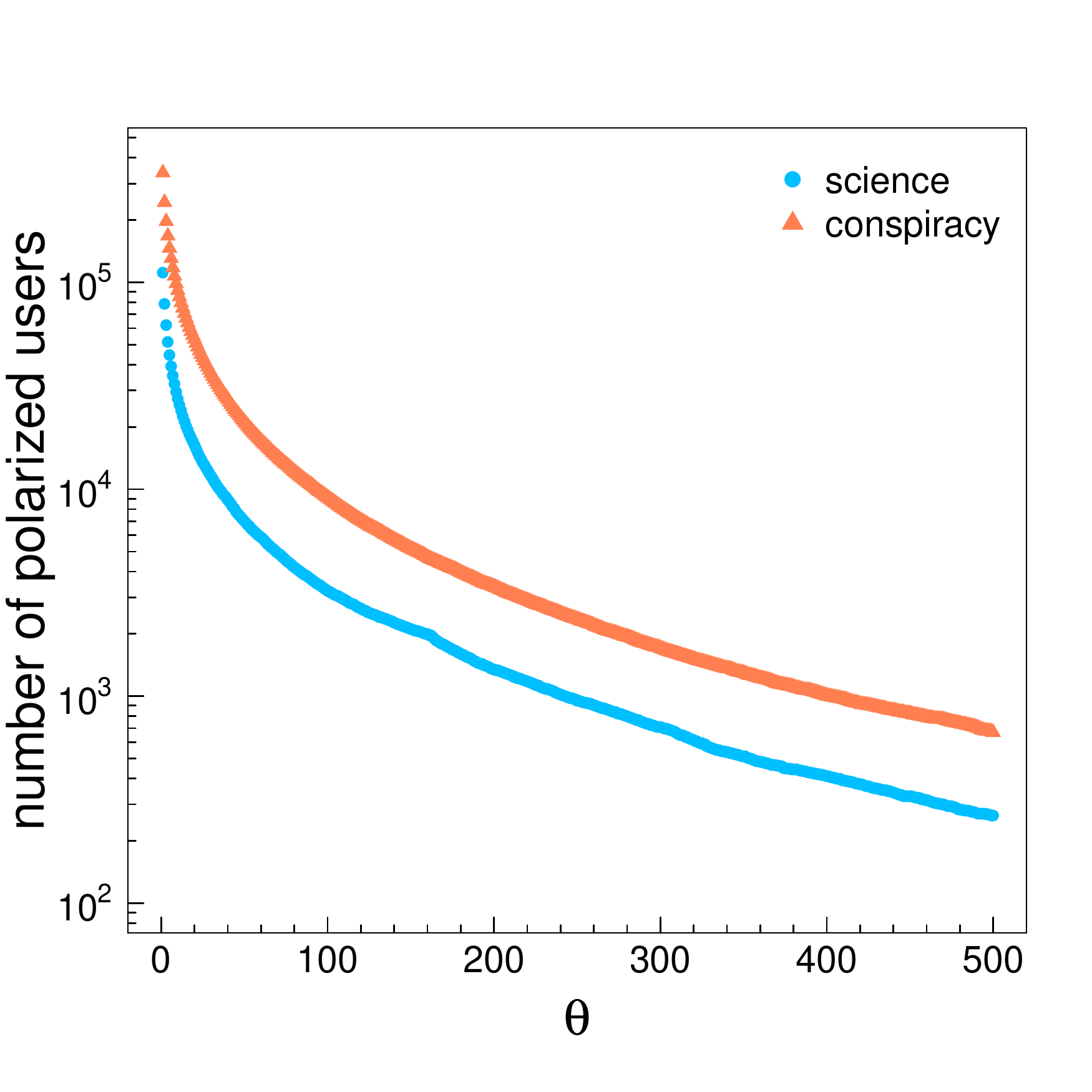}
\caption{\textbf{Users Engagement}. Number of polarized users as a function of the engagement degree $\theta$.}
\label{fig:num_engaged}
\end{figure}

We continue our analysis by addressing the relationship between the exposition to external information and the level of engagement of a user on his/her preferred narrative.
We assume as a good indicator of the user engagement within a given narrative the total number of likes posts. 
We test the response of polarized users of both categories to two different kind of information external to their narratives. 
We use information coming from {\em hoaxbusters} pages aiming at debunking and correcting the diffusion of false claims (mainly conspiracy theses) such as the link between vaccines and autism or the astonishing medical powers of soursop -- and {\em troll} pages intentionally posting satirical and demential imitations of conspiracy theses.  
Such a selection is peculiar for our analysis as it accounts for the polarization of debates between consumers of scientific news (supporters of rational thinking) and conspiracists that have been proved to be less rational and more prone to avoid scrutiny \cite{hogg2011extremism,finerumor,bauer1997resistance}.
In particular, we analyze how the activity (comments and likes) of polarized users on troll pages and debunking pages changes as a function of $\theta$ -- i.e. the engagement degree, intended as the number of likes of a polarized user in the category which he/she belongs to. In Figure \ref{fig:num_engaged} we show the number of polarized users as function of the threshold $\theta$. The two curves show a similar decreasing trend.

\begin{figure}[H]
 \centering
\subfigure[Trolls]
{\includegraphics[width=.47\textwidth]{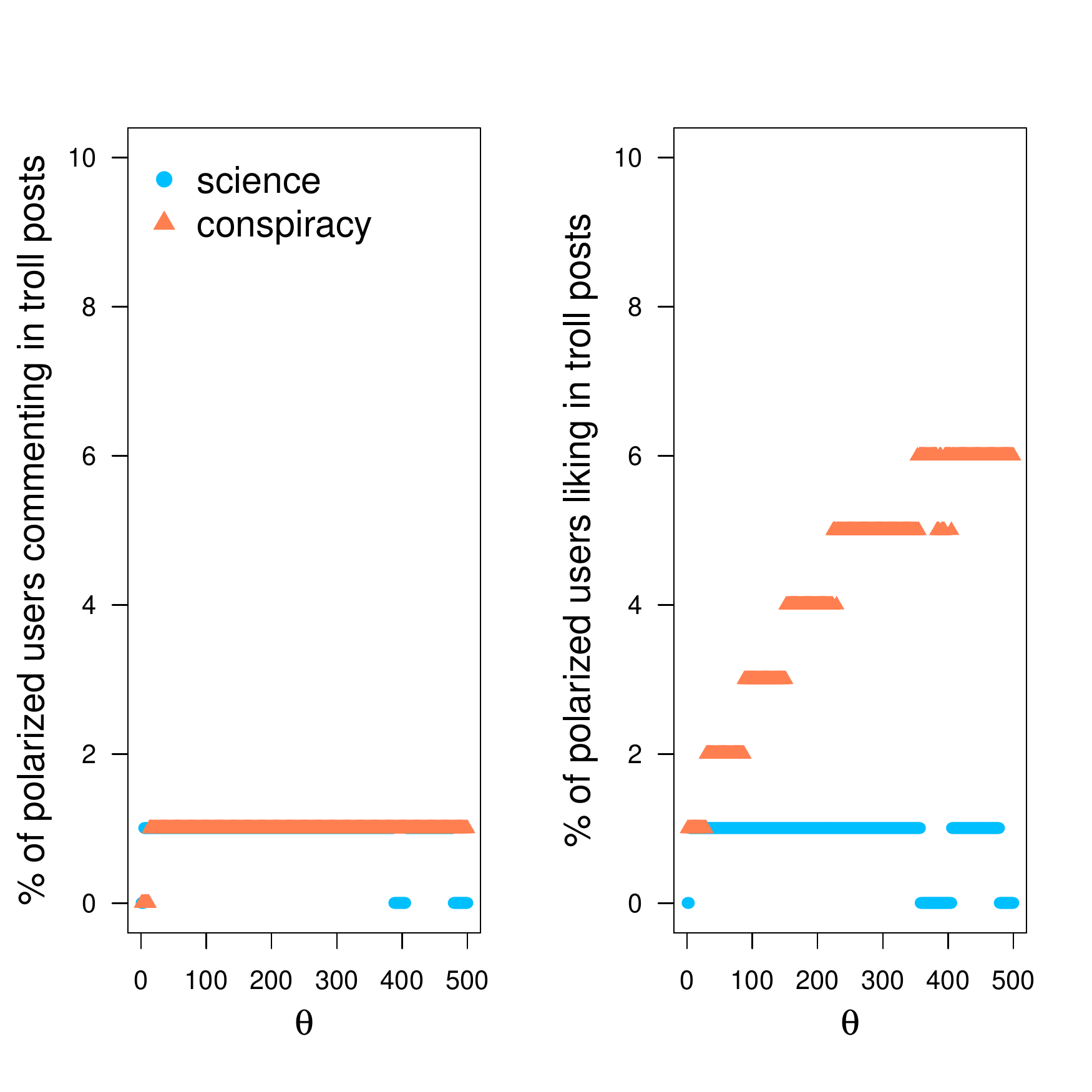} \label{fig:5}} 
\hspace{1mm}
\subfigure[Hoaxbusters]
{\includegraphics[width=.47\textwidth]{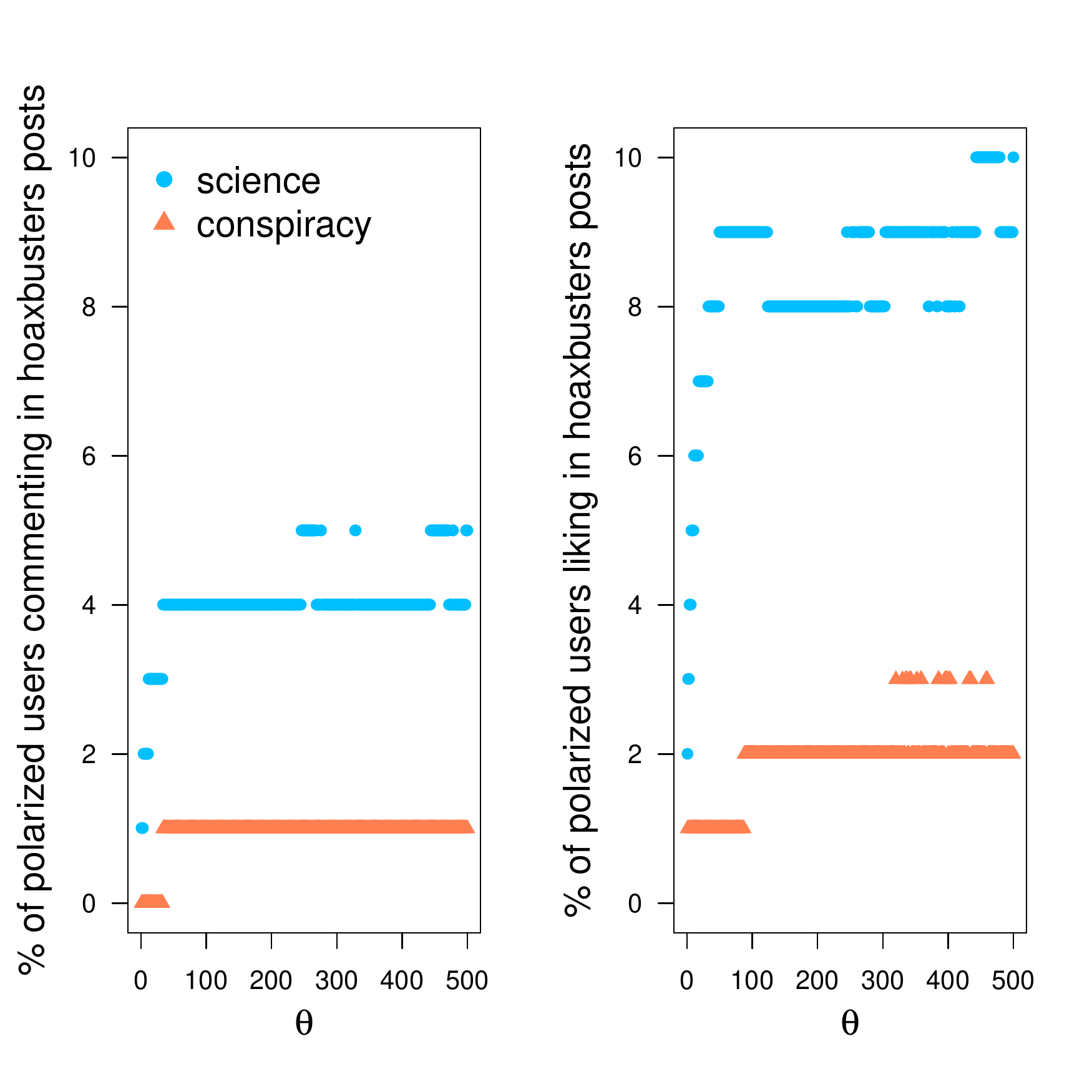} \label{fig:6}} 
\caption{\textbf{Users on external contents}. Users activity (likes and comments) as a function of the engagement degree $\theta$ of conspiracy and scientific news on troll (panel a) and hoaxbusters (panel b) posts.}
\label{fig:engagement}
\end{figure}

In Figure \ref{fig:engagement} we show the activity (number of likes and comments) of polarized users of scientific and conspiracy news on respectively,  4,502 debunking (panel a) and 4709 troll (panel b) information as a function of the user engagement $\theta$ on their related narrative.
On the one hand, consumers of scientific news are more active in liking and commenting debunking posts. On the other hand, consumers of conspiracist posts are more prone to like (and not to comment) satirical imitation of the story they are usually exposed to.
Such a trend of polarized users increases with their level of commitment and engagement (for the effective number of users refer to Figure \ref{fig:num_engaged}).

The results of Figure \ref{fig:engagement} suggest that conspiracists are interested in diffusing their stories; their tendency to avoid scrutiny allows for the mixing of conspiracy news and their satirical imitation, pointing out the high credulity level of consumers of conspiracy-related information.  
On the other hand, also polarized users of scientific news tend to like and comment information that are consistent with their narratives (debunking of unsubstantiated claims). 
Such results are a warning on the effectiveness of online debunking activities since they are mainly fruited by users of scientific pages and are not considered by consumers of conspiracy information.  
Coherently with \cite{Mocanu2014}, high levels of commitment in conspiracy theses decrease the level of interest in official and main stream information and increases the possibility to interact with unsubstantiated rumors. 

\subsection{Conspiracy news within online debunking and trolls}

\begin{figure}[H]
 \centering
       \includegraphics[width=0.95\textwidth]{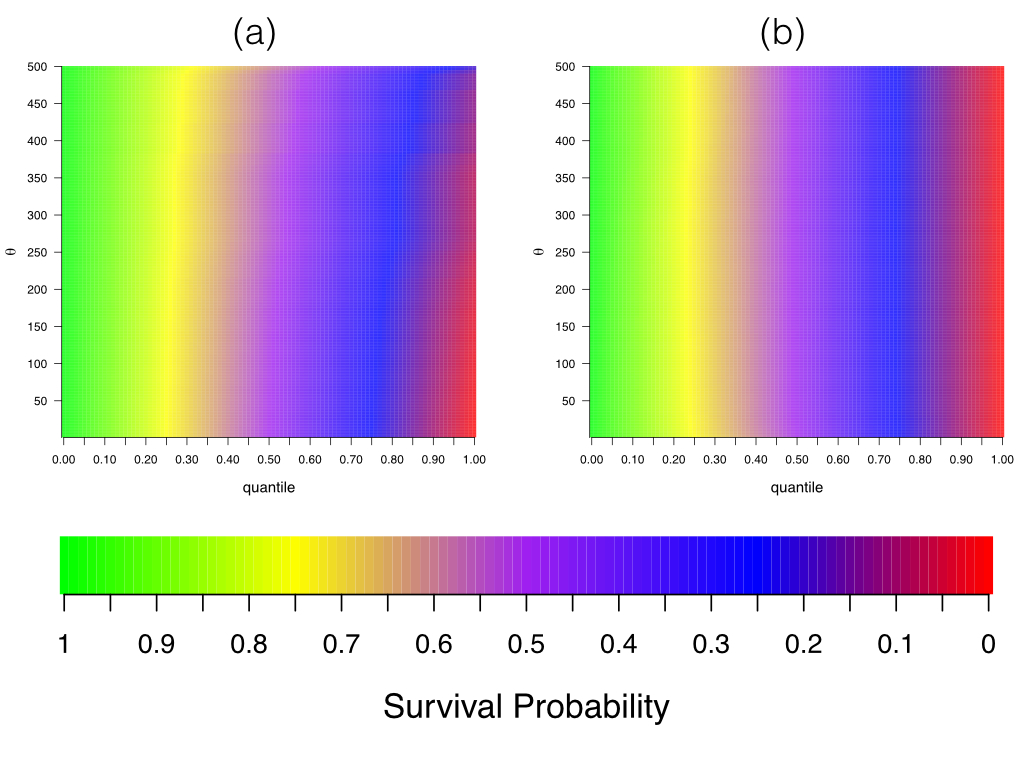} 
\caption{\textbf{Survival probability of conspiracists exposed to troll posts.} Heatmap of the quantile discretization of the survival probability distribution of conspiracy users against their level of engagement $\theta$ exposed (panel a) and not exposed (panel b) to satirical and demential imitation of the story they are usually exposed to.}
\label{fig:heat_comment_troll}
\end{figure}

\begin{figure}[H]
 \centering
       \includegraphics[width=0.6\textwidth]{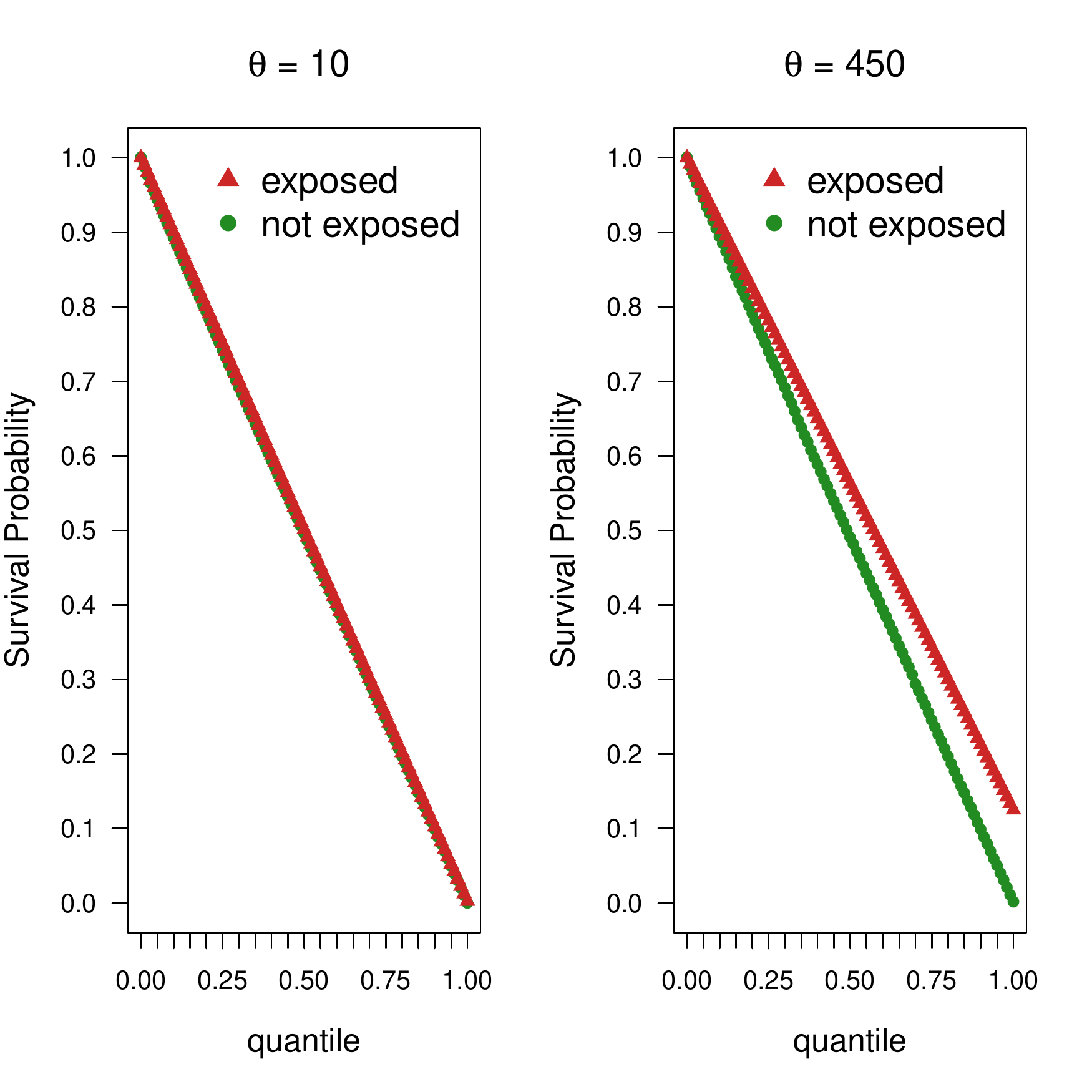} 
\caption{\textbf{Survival probability of conspiracists exposed to troll posts.} Quantile discretization of the survival probability for conspiracists exposed and not exposed to satirical and false information at different levels of engagement. We notice a nearly identical survival probability between exposed and not exposed when the level of engagement is low ($\theta = 10$), whereas when the level of engagement is high ($\theta = 450$) we find that the survival probability of conspiracists exposed to satirical and false information is progressively higher with respect to the one of those not exposed.}
\label{fig:trolls}
\end{figure}

Information consumption is driven by the user's system of beliefs. 
Hence, information aimed at contrasting the diffusion of unsubstantiated claims -- i.e. mainly targeting the debunking of conspiracy rumors -- are almost ignored by conspiracists. 
However, the interaction between conspiracists and debunkers posts might occur.
We test the effect on usual consumers of conspiracy news to the exposure to debunker posts. 
In particular, we want to understand if these posts are effective in changing the tendency of users to interact with unsubstantiated claims.
To do this we measure the survival probability of conspiracy users who commented (active interaction)  either posts from debunking pages or false information as a function of the level of user engagement $\theta$ in interacting with conspiracy news. 
Focusing on the persistence of polarized users on posts of their category we estimate their survival probability function. More precisely, we compute the probability that a user's lifetime -- i.e. the temporal distance between the first and the last like of the user in the category which he belongs to -- is greater than some specified temporal distance $t$.
Let  define the random variable $T$ with cumulative distribution function $F(t)$ on the interval $[0,\infty)$. Then the probability that a user's lifetime is not greater than a specific $t$ is given by the cumulative probability distribution $ F(t) = Pr(T \leq t). $
Hence, the survival function is the probability that a user will continue to like posts supporting the narrative in which he is polarized on beyond a given time $t$  given by $ S(t) = Pr(T > t) = 1 - F(t). $
To compute such a measure we use  the \emph{Kaplan Meier estimate} \cite{Kaplan58}.
Let $n_{t}$ denote the number of users that are still liking posts supporting the narratives in which they are polarized on, just before time $t$; and let $d_{t}$ denote the number of users that stop liking at time $t$. 
Then the estimated survival probability after time $t$ is $(n_{t} - d_{t})/n_{t}$.
Assuming that the times $t$ are independent, the Kaplan Meier estimate of the survival function at time $t$ is defined by $ \hat{S}(t) = \prod^{t}\left( \frac{n_{t} - d_{t}}{n_{t}} \right). $
Figure \ref{fig:heat_comment_troll} shows in panel (a) the quantile discretization, for different levels of engagement $\theta$, of the survival probability distribution of usual consumers of conspiracy news which interacted with troll posts; and in panel (b), as a control, the quantile discretization, for different levels of engagement $\theta$, of the survival probability distribution of polarized users not exposed to intentional false claims. Figure \ref{fig:trolls} shows the quantile discretization of the survival probability for conspiracists exposed and not exposed to satirical and false information at different levels of engagement, i.e. $\theta = 10$ and $\theta = 450$. For consumers of conspiracy news not exposed to troll memes, the probability to remain polarized is constant with the increase of their level of commitment.
Conversely, the more a user is engaged, the more a contact with a troll post will reinforce the probability to remain a polarized user in his category.

\begin{figure}[H]
 \centering
       \includegraphics[width=0.95\textwidth]{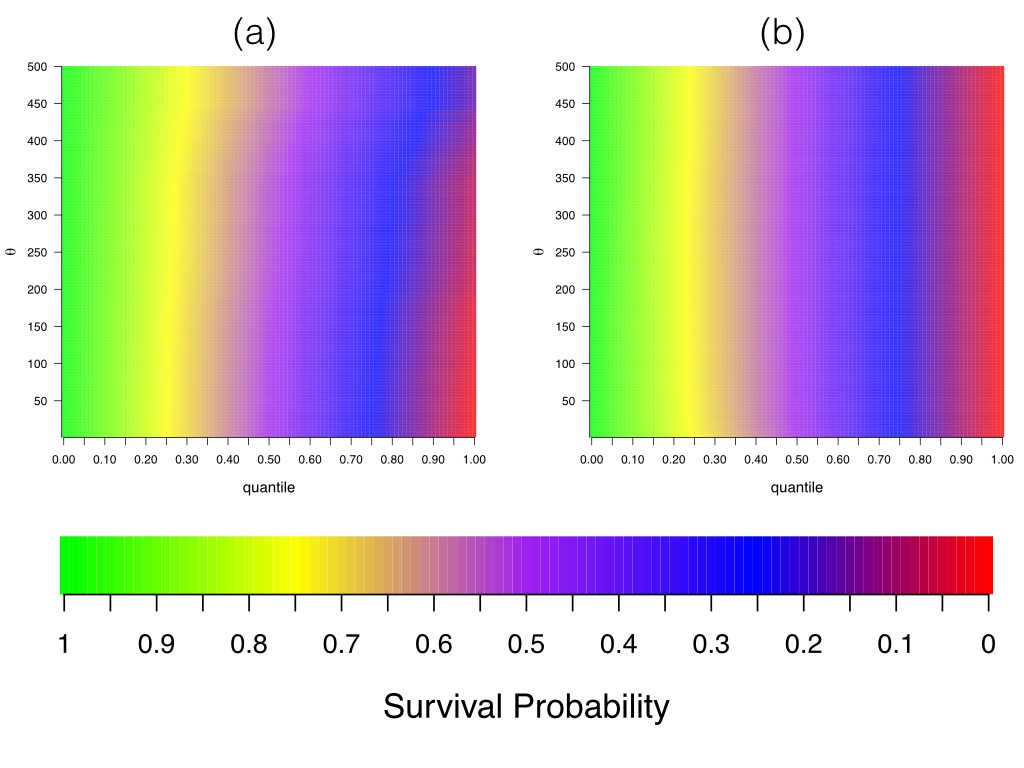} 
\caption{\textbf{Survival probability of conspiracists exposed to debunking posts}. Quantile discretization of the survival probability distribution of conspiracy users against their level of engagement $\theta$ exposed (panel a) and not exposed (panel b) to posts debunking conspiracy theses.}
\label{fig:heat_comment_hoax}
\end{figure}

\begin{figure}[H]
 \centering
       \includegraphics[width=0.6\textwidth]{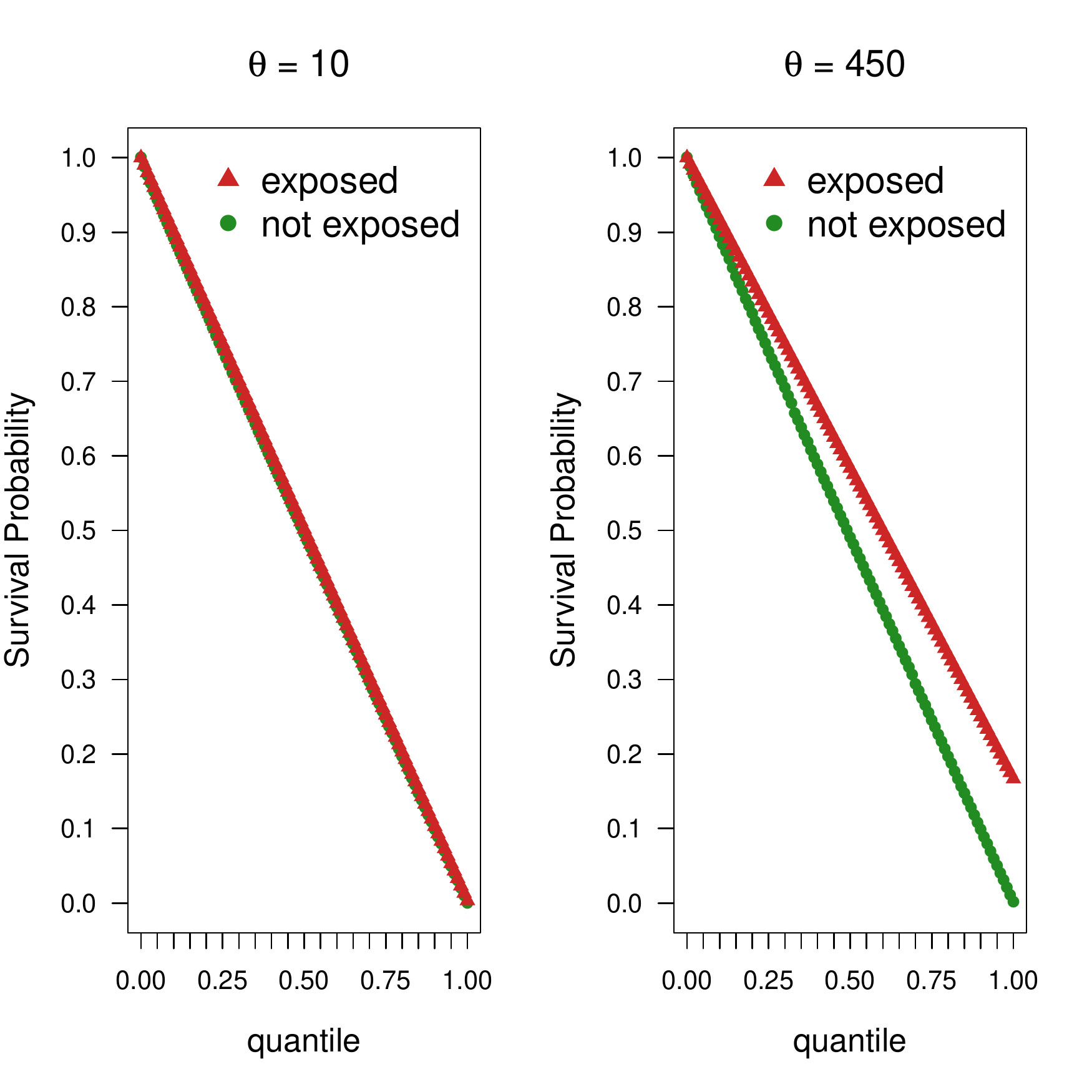}
\caption{\textbf{Survival probability of conspiracists exposed to debunking posts.} Quantile discretization of the survival probability for conspiracists exposed and not exposed to posts debunking conspiracy theses at different levels of engagement. We notice a nearly identical survival probability between exposed and not exposed when the level of engagement is low ($\theta = 10$), whereas when the level of engagement is high ($\theta = 450$) we find that the survival probability of conspiracists exposed to debunking posts is progressively higher than the not exposed.}
\label{fig:hoaxbusters}
\end{figure}

Similar results hold for the reaction to information having the goal to persuade users of the unsubstantiated nature of conspiracy thesis. Figure \ref{fig:heat_comment_hoax} shows the quantile discretization of the survival probability distribution for increasing level of users engagement $\theta$ of usual consumers of conspiracy news exposed (panel a) and not exposed (panel b) to debunking memes. Figure \ref{fig:hoaxbusters} shows the quantile discretization of the survival probability for conspiracists exposed and not exposed to debunking posts at different levels of engagement, i.e. $\theta = 10$ and $\theta = 450$. For consumers of conspiracy news not exposed to debunking memes, the probability to remain polarized is constant with the increase of their level of commitment. Conversely, the more a user is engaged, the more a contact with a debunking post will reinforce the probability to remain a polarized user in his category.

These results suggest that the more a user is committed in consuming conspiracy related information, the more the injection of either satirical  or debunking posts will increase the probability to continue consuming conspiracy stories.
Either contrasting or making fool of biased narratives might create an unwanted reinforcement and burst on the diffusion of conspiracy theses and unsubstantiated rumors fostering the formation of biased beliefs.

\section{Conclusions and future works}

Conspiracy theories as alternative explanations to complex phenomena (e.g. globalization) find on the Web a natural medium for their dissemination and, not rarely, are used as argumentation for policy making. 
We measure the effect of the exposition to intentional false claims (satirical and demential version of conspiracy theses) and to debunk memes (information aiming at correcting the diffusion of false claims) for increasing level of user commitment on the preferred narrative. Our results show the exposure to both claims and debunking might reinforce the probability to interact again with conspiracist stories.
As a benchmark,  we use information coming from debunking pages, namely hoaxbusters -- i.e. pages aiming at correcting the diffusion of false claims such as the link between vaccines and autism, or the global warming caused by chemtrails -- and troll pages intentionally posting satirical and demential imitation of conspiracy theses.
In the next future we aim at exploring the potential implication of the friendship network structure in the creation, diffusion and reinforcement of narratives.

\section*{Acknowledgments}
Funding for this work was provided by EU FET project MULTIPLEX nr. 317532 and SIMPOL nr. 610704. The funders had no role in study design, data collection and analysis, decision to publish, or preparation of the manuscript.
We want to thank "Protesi di Protesi di Complotto", "Scientificast", "Simply Humans" and "Semplicemente me" page staff.
Additional special thanks go to Sandro Forgione, Salvatore Previti, Fabio Petroni, La Titta, and Elio Gabalo for precious suggestions and discussions.
\bibliographystyle{unsrt}
\bibliography{trolling}

\end{document}